# Anisotropic Upper Critical Field, Seebeck and Nernst Coefficient in $Nb_{0.20}Bi_2Se_3$ Topological Superconductor

Shailja Sharma, C.S. Yadav*

*School of Basic Sciences, Indian Institute of Technology Mandi, Mandi-175005 (H.P.) India*
*Email: shekhar@iitmandi.ac.in

We present the magneto-transport and the thermoelectric (Seebeck and Nernst coefficient) studies of the Nb-doped $Bi_2Se_3$ topological superconductor. The angle-dependent magnetoresistance study highlights the anisotropy in the upper critical field ($H_{c2}$) with the anisotropy parameter $\Gamma \sim 1.2$. We observed a gradual decrease in low-temperature Hall resistivity on the application of magnetic field like any conventional superconductor, instead of the finite Hall resistivity which was linked to the chiral superconducting phase. The estimated value of the carrier concentration ($\sim 10^{19}$ cm$^{-3}$) for $Nb_{0.2}Bi_2Se_3$ is one order larger than for $Bi_2Se_3$. Doping of Nb shows a significant decrease in the Seebeck coefficient value and the estimated Fermi temperature of the three-dimensional Fermi surface at the centre of Brillouin zone in the zero-temperature limit enhances by ~4 times in comparison to pristine $Bi_2Se_3$. We have observed a large value (~2.3 μV K$^{-1}$ T$^{-1}$) of Nernst coefficient for $Bi_2Se_3$ at room temperature which decreases with Nb doping ( ~0.5 μV K$^{-1}$ T$^{-1}$).

**Introduction:**

Three-dimensional topological insulators (TIs) are characterized by their non-trivial surface states, where electrons have spin-momentum locking with time-reversal symmetry.[1] $Bi_2Se_3$ has been identified as a promising TI material for the realization of topological superconductivity.[2] Superconductivity can be realised by chemical doping, by applying pressure and by proximity effect.[3-5] These topological superconductors (TSCs) are characterized by bulk superconducting gap with gapless conducting surface states.[2] Unlike the TIs, where the surface states consist of Dirac fermions, the surface states in TSCs consist of Majorana fermions, which are their own antiparticles.[6] The TSCs possess gapless excitations on the boundary due to Bogoliubov quasiparticles which are the coherent superpositions of electrons and holes.[7] The experiments have revealed the unique features of zero-bias conductance peak (ZBCP) and the Majorana zero modes (MZMs) in the TSCs.[2, 8, 9] Moreover, angle resolved photoelectron spectroscopy (ARPES) measurements in TSCs showed the existence of non–trivial surface states in normal states.[10-12]

Recent experiments reveal that carrier-doped $Bi_2Se_3$ is a nematic TSC, with two-component $E_u$ order parameter that belongs to crystal point group $D_{3d}$, which spontaneously break three-fold rotational symmetry, leading to a nematic order.[13-15]. Nematic superconductivity has been observed in Cu, Sr, and Nb doped $Bi_2Se_3$ superconductors[16] below superconducting transition temperature ($T_c$) in the angle dependent measurements of spin susceptibility, heat capacity, magnetoresistance, and magnetic torque.[12, 16-22] The study of temperature dependence of London penetration depth indicate the presence of symmetry-protected point nodes consistent with $E_u$ odd-parity pairing.[23] Further, the penetration depth study on the proton irradiated sample showed odd frequency pairing for superconductivity where order parameter has symmetry protected nodes.[24] These results suggest the realization of topological superconductivity in the nonmagnetic disordered system.[24] Asaba *et al.* investigated the nematicity in hysteresis through torque magnetometry in $Nb_xBi_2Se_3$ and also showed nodeless superconducting gap, which is consistent with odd parity *p*-wave pairing.[20] The quantum oscillation results on Nb-doped $Bi_2Se_3$ have shown the multi orbit nature of the electronic state which is in contrast to one bulk Fermi pocket in $Bi_2Se_3$, $Cu_xBi_2Se_3$ and $Sr_xBi_2Se_3$.[25] In a recent study Cho *et al.* has shown the presence of superconducting-fluctuation-induced nematic order above $T_c$.[26] Therefore, superconductivity in $Nb_xBi_2Se_3$ compound offers rich physics from the point of view of nematicity in different quantities as well as electronic transport discussed via quantum oscillations. The Seebeck and Nernst effect in $Nb_{0.20}Bi_2Se_3$ TSC has remain unexplored. Nernst effect provides an insight to understand the electronic properties of exotic topological materials.

In this article, we present detailed angle-dependent magneto-transport and thermoelectric studies on the good quality single crystalline $Nb_{0.20}Bi_2Se_3$ superconductor. The methodology for single crystal growth and experimental techniques used for the study are shown in the supplementary material. We have discussed the anisotropic upper critical field ($H_{c2}$) through angle-dependent magneto-transport study in the superconducting state. Our studies rules out the presence of spontaneous magnetization and the Hall signal in superconducting phase, which has been linked to the chiral superconductivity in the compound. Furthermore, the thermoelectric properties of $Nb_{0.2}Bi_2Se_3$ are discussed through temperature-dependent Seebeck and Nernst coefficient data in the zero-temperature limit within the single-band picture for Fermi liquid metals, which are compared with the results on pristine $Bi_2Se_3$.

**Results and discussion:**

*Magneto-transport studies:* The longitudinal resistivity $\rho_{xx}(T)$ of $Nb_{0.20}Bi_2Se_3$ shown in Fig. 1(a) follows metallic behaviour down to 3.5 K, below which it shows superconductivity in agreement with the literature.[25] Inset shows the temperature-dependent magnetization ($M$) in both zero-field cooled (ZFC) and field cooled (FC) measured at $H$ = 10 Oe. The magnetisation measurements were performed in a configuration when H // *ab* plane to minimize the effect of demagnetization factor.

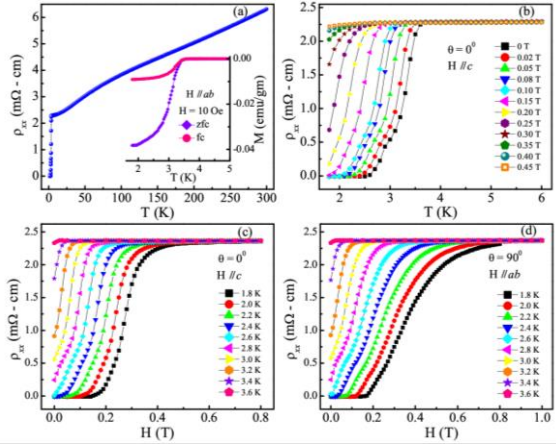

Figure 1: (a) The $\rho_{xx}(T)$ measured at $H$ = 0 Oe, showing the superconducting transition at 3.5 K, Inset shows the magnetisation, $M$ ($T$) measured at $H$ = 10 Oe. (b) The $\rho_{xx}(T)$ measured at $H$ = 0 – 0.45 T, (c) and (d) show the plots for $\rho_{xx}(H)$ measured at $T$ = 0 - 3.6 K for $H$ // $c$ -axis and $H$ // $ab$ plane, respectively.

The low temperature $\rho_{xx}(T)$ measured at different fields is presented in Fig. 1(b) for $H$ // $c$-axis of the crystal. With increasing magnetic fields, $T_c$ shifts to lower temperature and gradually suppressed to zero. Further, magnetic field dependence of resistivity $\rho_{xx}(H)$ at different temperatures below $T_c$ is shown for different orientation of the sample, $H$ // $c$-axis and $H$ // $ab$- plane in Fig. 1(c) and Fig. 1(d), respectively. It is to note that the superconducting transition gets more broaden for $H$ // $ab$- plane in comparison to $H$ // $c$-axis. Figure 2(a) shows the $M$-$H$ loop with H // $ab$-plane at 1.8 K revealing the type-II superconducting behaviour without any magnetic Nb contribution, in agreement with the hysteresis loop as discussed by Shen *et al.*[27] Figure 2 (b) shows the schematic for the angle-magnetic field dependent resistivity measurements, where $\theta = 0°$ corresponds to the geometry when $H$ // $c$-axis and $\theta = 90°$ when $H$ // $ab$- plane, and current is always perpendicular to the field direction. We have extracted the upper critical field ($H_{c2}$) from the above resistivity curves as shown in Fig. 2(c). The $H_{c2}$ values at different temperatures were obtained from the fields at which the resistivity of the sample has reached to 50 % of the normal state resistivity values at different temperatures. Applying the Werthamer-Helfand-Hohenberg (WHH) formula, which describes the behaviour of upper critical field in conventional type-II Bardeen-Cooper-Schrieffer (BCS) superconductors, we extracted $H_{c2}(0)$ using[28] $\mu_0 H_{c2}^{orb}(0) = -0.693 \left(\frac{d\mu_0 H_{c2}}{dT}\right)_{T_c} T_c$ , where the slope $|dH_{c2}(T)/dT|$ can be obtained from the linear fitting of the curves in graph. The value of zero-temperature orbital limited upper critical field, $\mu_0 H_{c2}^{orb}(0)$, along the $c$-axis and parallel to $ab$-plane, were calculated to be 0.42 T and 0.55 T, respectively. Using $H^P(0) = \Delta/\sqrt{2}\mu_B$ and $\Delta = 1.76 k_B T_c$ we evaluate the Pauli limiting field $H^P(0)$ = 6.1 T for H // $c$-axis. When both the orbital and spin limiting fields are present, the critical field is given as $H_{c2}(0) = H_{c2}^{orb}(0)/\sqrt{1 + \alpha^2}$, with the Maki parameter, $\alpha = \sqrt{2} H_{c2}^{orb}(0)/H^P(0)$.[28, 29] The $\alpha$ = 0.098(0.13) and $H_{c2}(0)$ = 0.42 T (0.55 T) for H // $c$-axis ($ab$-plane). Since $\alpha$ is less than one hence the Flude-Ferrell-Larkin-Ovchinnikov (FFLO) state is not favorable.[30] Since $H_{c2}^{orb}(0) \sim H_{c2}(0) < H^P(0)$, superconductivity in $Nb_{0.20}Bi_2Se_3$ is Pauli limited, as earlier reported for Sr and Cu-doped $Bi_2Se_3$.[22, 31]

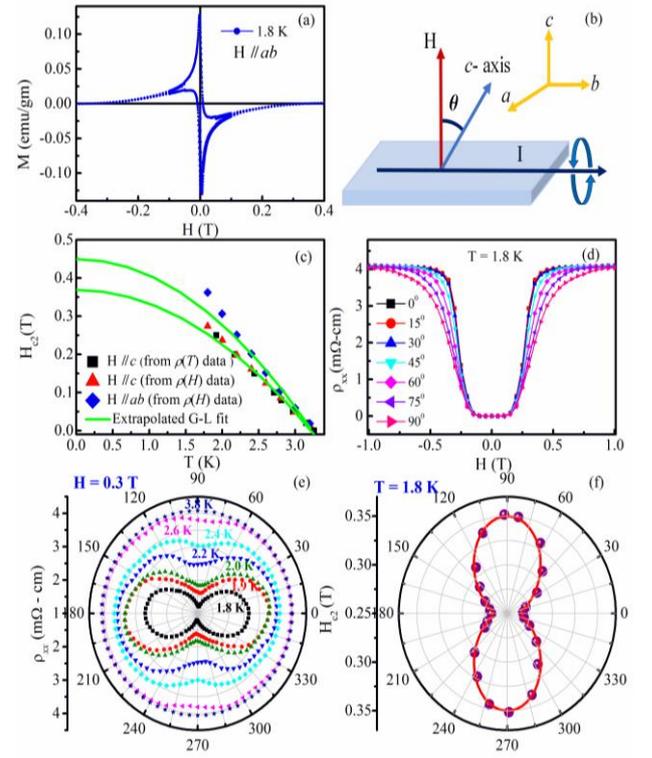

Figure 2:(a) $M(H)$ at 1.8 K showing superconducting hysteresis loop. (b) Schematic of angular-magneto-transport measurement geometry: Angle ($\theta = 0^0$) is when magnetic field is parallel to $c$ - axis ($H$ // $c$), and $\theta = 90^0$ is when magnetic field is parallel to $ab$ – plane ($H$ // $ab$). (c) $H$–$T$ phase diagram along with the extrapolated Ginzburg-Landau upper critical field. (d) The $\rho_{xx}(H)$ measured at T = 1.8 K at different angles, showing the anisotropy in $\rho_{xx}(H)$. (e) Polar plot showing $\rho_{xx}$ ($T$, $\theta$) at H = 0.3 T below $H_{c2}$ measured between $T$ = 1.8 K and 3.8 K. (f) Polar diagram showing anisotropy of $H_{c2}$ at T = 1.8 K obtained from fig(d), red curve shows the anisotropic Ginzburg-Landau fit (eq.1).

The $H_{c2}$ vs $T$ data as shown in Fig. 2(c), were fitted using equation $H_{c2}(T) = H_{c2}(0)\left[1 - \left(\frac{T}{T_c}\right)^2\right]$ from the Ginzburg-Landau (GL) theory, where $H_{c2}(0)$ is the zero-temperature upper critical field and $T_c$ is 3.2 K. The effective upper critical field for anisotropic superconductor varies between two orientations depending upon the superconducting anisotropic ratio, $\Gamma = H_{c2}^{\parallel ab}(0)/H_{c2}^{\parallel c}(0)$ is ~ 1.2. Similar values have been discussed for Cu, Nb, Sr doped $Bi_2Se_3$.[22, 23, 31] The anisotropy in $H_{c2}$ is attributed to the layered crystal structure, which has been ascribed to the dimensional crossover in the literature.[32, 33] The superconducting coherence length can be extracted from the following expressions: $H_{c2}^{\parallel c} = \frac{\Phi_0}{2\pi\xi_{ab}^2}$, $H_{c2}^{\parallel ab} = \frac{\Phi_0}{2\pi\xi_{ab}\xi_c}$ ; where $\Phi_0 = 2.07 \times 10^{-7}$ Oe-cm$^2$ is the magnetic quantum flux, and $\xi_{ab}$ and $\xi_c$ are the superconducting coherence length in the *ab* plane and along the *c*-axis, respectively. The superconducting coherence lengths $\xi_{ab}(0)$ and $\xi_c(0)$ at 0 K, were estimated to be ~30 nm and ~25 nm, respectively. The carrier density calculated using the one-band Drude model from the Hall data at 5 K, and assuming a spherical Fermi surface where $k_F = (3\pi^2 n)^{1/3}$, mean free path, $l$ can be estimated from the relation $l = \hbar k_F / \rho_0 n e^2$. With $n = 3.33\times10^{19}$ cm$^{-3}$ and $\rho_0 = 2.29\times10^{-3}$ $\Omega$-cm ($T$ = 5 K), we calculate $k_F = 8.47 \times 10^6$ cm$^{-1}$ and $l$ = 4.5 nm, which is less than the coherence length. Comparing the mean free path, $l$ and coherence length, $\xi_{ab}$, $\xi_c$, we find that for our sample, $l < \xi_{ab}$, $\xi_c$, which is not in the clean limit. Figure 2(e) shows the polar plot for the angular dependence of resistivity measured in fixed applied magnetic field $H$ = 0.3 T (below the $H_{c2}$ value) in the superconducting regime at different temperatures ($T$ = 1.8, 1.9, 2.0, 2.2, 2.4, 2.6 and 3.8 K). The angular dependence of resistivity reflects the two-fold anisotropy in superconducting state. We see that normal state resistance ($T$ = 3.8 K) does not show any variation for different angles. Therefore, the two-fold rotational symmetry in SC state becomes rotationally isotropic in normal state.

Furthermore, we measured $\rho_{xx}(H)$ at $\theta$ = 0, 15, 30, 45, 60, 75, and 90° with respect to *c*-axis of sample at ± 1T field in superconducting state at $T$ = 1.8 K ((see Fig. 2(d)). It is to mention here that the current is always kept perpendicular to the magnetic field directions. We find that the $H_{c2}$ value increases as magnetic field gets parallel to *ab*-plane of sample ($\theta$ = 90°). The extracted $H_{c2}$ values are plotted with different angles in polar plot in Fig. 2(f). The graph clearly depicts the two-fold anisotropy of $H_{c2}$ in superconducting state. This can be understood according to anisotropic Ginzburg-Landau theory, effective mass anisotropy leads to the anisotropy of the upper critical field.[34] The upper critical field at an angle $\theta$ is given by

$$H_{c2}^{GL}(\theta) = \frac{H_{c2}^{\parallel c}(0)}{\sqrt{\cos^2(\theta) + \Gamma^{-2}\sin^2(\theta)}} \quad (1)$$

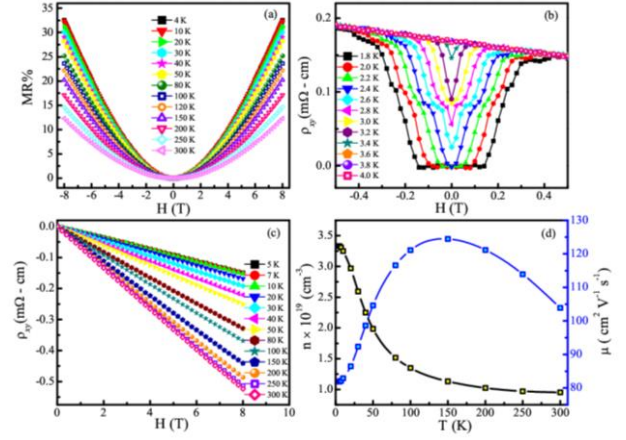

Figure 3: (a) MR at various temperatures (4-300 K). (b) The $\rho_{xy}(H)$ at low fields ($H$ = ± 0.5 T) and low temperature, below T = 3.6 K. (c) The $\rho_{xy}(H)$ for T = 5 – 300 K, and (d) Temperature variation of carrier density (*n*), and mobility (*μ*).

where $\theta$ is the angle of the applied magnetic field measured from the direction perpendicular to the plane of the layers *i.e.* along the *c*-axis, with $\theta = \pi/2$ is when field is parallel to the plane of layers, *i.e.* along the *ab*-plane, and $\Gamma = \sqrt{m_c^*/m_{ab}^*} = \sqrt{M/m}$, ($M>m$) anisotropy between the effective masses of the quasiparticles along the *c*-axis and the *ab* plane, which is also related to the resistivity and the upper critical field anisotropy, $\Gamma = \rho_c/\rho_{ab}$, $\Gamma = H_{c2}^{\parallel ab}(0)/H_{c2}^{\parallel c}(0)$ respectively.[32] We fit the data with the formula (eq.(1)) that describes the two-fold anisotropy of layered superconductors, with $\Gamma$ = 1.3, and $H_{c2}(0)$ = 0.26 T. Therefore, effect of crystal anisotropy is evident in two-fold anisotropy of both resistivity and upper critical field data in superconducting state.

Further, we performed magnetoresistance measurements at various temperatures in normal state (above 4 K), in magnetic fields ± 8 T as illustrated in Fig. 3(a). The percentage change in MR is defined as $MR\% = \frac{\rho(H) - \rho(H=0)}{\rho(H=0)} \times 100$ %. The $Nb_{0.2}Bi_2Se_3$ is found to exhibit $H^2$ dependence at low fields and non-saturating linear behaviour at high fields, similar to the parent compound $Bi_2Se_3$ with a maximum value of ~ 35 % at $T$ = 4 K which is comparable to that of $Bi_2Se_3$.[35] As observed from the plots that quadratic field dependence of MR increases with increase in temperature. The quadratic dependence of MR is attributed to the deflection of charge carriers due to the Lorentz force under the applied magnetic fields, which gets saturated at high fields, followed by linear magnetoresistance (LMR). LMR is usually observed in topological materials, explained using classical PL model[36] and quantum model by Abrikosov.[37] The LMR at low temperatures seems to originate from the linear energy dispersion, as observed for the gapless topological surface states observed in (Bi, Sb)$_2$(Se, Te)$_3$, Dirac and Weyl semimetals and other topological materials.[38-40]

The Hall resistivity measured in low temperature (1.8-3.6 K) and low fields (±0.5 T), is presented in Fig. 3 (b). The observed $\rho_{xy}(H)$ behaviour is a typical of a BCS superconductor, and the $\rho_{xy}(H)$ drops to zero in superconducting state. It is reported that Nb doped $Bi_2Se_3$ spontaneously breaks time reversal-symmetry due to finite magnetisation and observation of Hall resistivity in zero-field below $T_c$.[41, 42] Our Hall effect studies are in sharp contrast with above studies, since the Hall effect is zero in superconducting state which is true for a superconductor. This is also supported by $M$-$H$ isotherm measured at $T = 1.8$ K plot as depicted in Fig. 2(a), which does not show the presence of any additional magnetic contribution in superconducting state. These results are contrary to the claims of the time-reversal symmetry breaking in superconducting states.

The $\rho_{xy}(H)$ measured at $T = 5 - 300$ K and $H = 0 - 8$ T is presented in Fig. 3 (c). The $\rho_{xy}(H)$ is linear in the complete field range with negative slope implying electrons as dominant charge carriers in this system. The carrier density ($n$) and Hall mobility ($\mu$) are calculated based upon the one-band Drude model and the temperature dependence is shown in Fig. 3 (d). The carrier density and mobility lie within the range (1 - 3.5)×$10^{19}$ cm$^{-3}$ and (80 – 125) cm$^2$ V$^{-1}$ s$^{-1}$, respectively. The carrier density is one order larger compared to the parent $Bi_2Se_3$,[35] but comparable to that reported in literature for doped $Bi_2Se_3$.[22, 43, 44]

*Thermoelectric studies:*

Temperature dependence of Seebeck coefficient ($S_{xx} = E_x/|\nabla T_x|$) and Nernst coefficient ($\nu = S_{xy}/H$; $S_{xy} = E_y/|\nabla T_x|$) are presented in Fig. 4 for undoped $Bi_2Se_3$ and $Nb_{0.20}Bi_2Se_3$ single crystal. Inset shows the schematic for the Seebeck and Nernst effect measurement, where thermal gradient was applied along the *ab*-plane and the magnetic field was oriented along the *c*-axis. The negative value of $S_{xx}$ shows *n*-type conduction for both compounds with the room temperature values of $S_{xx} \sim -140$ μV/K and $\sim -38$ μV/K for $Bi_2Se_3$ and $Nb_{0.20}Bi_2Se_3$ respectively. A clear superconducting transition for $Nb_{0.20}Bi_2Se_3$ could not be observed as $S(T)$ value becomes very close to zero in low-temperature limit ($S = -0.04$ μV/K at $T = 4$ K). Generally, $S_{xx}$ in metals is a sum of diffusion term ($S_{diff}$) and the phonon-drag term ($S_{drag}$). Within the Fermi liquid picture and in the absence of phonon drag, $S_{xx}(T)$ shows linear temperature dependence given by the Mott expression[45],

$$S_{xx}/T = \pm \frac{\pi^2}{2}\frac{k_B}{e}\frac{1}{T_F} = \pm \frac{\pi^2}{3}\frac{k_B^2}{e}\frac{N(\epsilon_F)}{n}$$

where $k_B$ is Boltzmann's constant, e is electron charge, n is the carrier density, and $T_F$ is the Fermi temperature. The density of states $N(\epsilon_F)$ is related to the Fermi energy as $N(\epsilon_F) = 3n/2k_BT_F$. Figure 4(b) presents the the temperature dependence of $S_{xx}/T$ for both the compounds. The slope of $S_{xx}/T$ vs. T curve is inversely proportional to $T_F$. The zero-temperature extrapolated value of $S_{xx}/T$ for $Bi_2Se_3$ and $Nb_{0.20}Bi_2Se_3$ are - 0.83 μV/K$^2$ and - 0.16 μV/K$^2$, that corresponds to $T_F$ values of 516 K and 2536 K, respectively. The data is consistent with the large number of n-type charge carriers observed for $Nb_{0.20}Bi_2Se_3$ ($10^{19}$ cm$^{-3}$) than parent $Bi_2Se_3$ ($10^{18}$ cm$^{-3}$). The obtained $T_F$ values for $Nb_{0.20}Bi_2Se_3$ are similar to those for other TSCs such as $Cu_xBi_2Se_3$[46] and $Sr_xBi_2Se_3$[22].

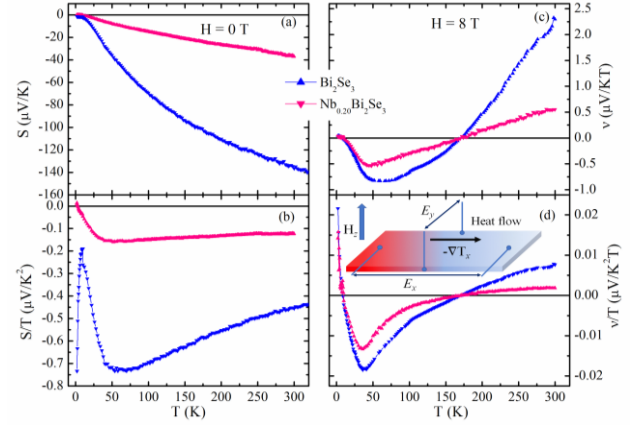

Figure 4: (a), (b) Temperature dependence of Seebeck coefficient, $S_{xx}$ and $S_{xx}/T$, in zero-field, respectively, and (c), (d) Temperature dependence of Nernst coefficient, $\nu$ and $\nu/T$ at field 8 T, respectively. Inset shows the schematic for Seebeck and Nernst effect measurement.

The value of Sommerfeld coefficient ($\gamma$) and Debye temperature ($\Theta_D$) were estimated from low temperature specific heat data (not shown here). The Sommerfeld coefficient $\gamma$ is given by $\gamma = \frac{\pi^2}{2}k_B\frac{n}{T_F} = \frac{\pi^2}{3}k_B^2 N(\epsilon_F)$.[45] We find that $S_{xx}/T$ and $\gamma$ values decrease upon Nb doping in $Bi_2Se_3$, reflecting a change in Fermi surface. With a simple Debye model to the phonon contribution, Debye temperature ($\Theta_D$) was calculated using $\Theta_D = \left(\frac{12\pi^4}{5\beta}nR\right)^{1/3}$, where $R = 8.314$ J mol$^{-1}$ K$^{-1}$ is the ideal gas constant and *n* is number of atoms for $Bi_2Se_3$. The density of states at the Fermi energy is calculated as $N(\epsilon_F) = \frac{3\gamma}{\pi^2 k_B^2}$. We find that the density of states decreases with Nb doping.

Considering the BCS theory of superconductivity, the electron-phonon coupling strength ($\lambda_{ph}$) can be estimated from the McMillan formula[47] $\lambda_{ph} = \frac{1.04+\mu^*\ln(\frac{\Theta}{1.45T_C})}{(1-0.62\mu^*)\ln(\frac{\Theta}{1.45T_C})-1.04}$, where $\mu^*$ is the coulomb coupling parameter empirically assigned to 0.13. The $\lambda_{ph}$ was estimated to be 0.7, indicating a moderate coupling strength (BCS weak coupling case: $\lambda_{ph} \ll 1$; strong-coupling case: $\lambda_{ph} > 1$). Using $S_{xx}/T$ and $\gamma$, we derived dimensionless parameter $q$, given by $q = \frac{S}{T}\frac{N_{Av}e}{\gamma}$, where $N_{Av}$ is the Avogadro number[45]. The obtained $q$ values are shown in table 1. Earlier, large value of $q$ has been observed for 'kondo insulator'.[45] Thus, we find that $S_{xx}/T$ and $\gamma$ values decrease upon Nb doping in Bi$_2$Se$_3$, which reflects a change in Fermi surface. Additionally, the ratio of $T_c$ to $T_F$ can be used to get information about the electronic correlation strength in superconductors.[48] The ratio $T_c/T_F \sim 0.0013$, implies a weakly correlated superconductor.[49] Assuming single band transport and a spherical Fermi surface, we have calculated the Fermi wave vector ($k_F$), effective mass ($m^*$) and Fermi velocity ($v_F$), using $k_B T_F = \hbar k_F^2 / 2m^*$ and $\hbar k_F = m^* v_F$, which are shown in table I.

Table-I: Parameters obtained from the value of $S/T$ in zero temperature limit, electron mobility, Fermi temperature, and other parameters within the single band picture of Fermi liquid.

| Sample | Bi$_2$Se$_3$ | Nb$_{0.20}$Bi$_2$Se$_3$ |
|---|---|---|
| $S^m/T$ (μV K$^{-2}$) | -0.73 | 0.0098 |
| $v^m/T$ (μV K$^{-2}$T$^{-1}$) | 0.022 | 0.014 |
| $T_F$ (K) | 516 | 2536 |
| $\mu$ (cm$^2$ V$^{-1}$s$^{-1}$) | 3844 | 82 |
| $\Theta_D$ (K) | 150 | 144 |
| $\gamma$ (mJ mol$^{-1}$ K$^{-2}$) | 1.76 | 0.24 |
| $N(\epsilon_F)$ (states eV$^{-1}$ per f.u.) | 0.75 | 0.10 |
| $q = \frac{S}{T}\frac{N_{Av}e}{\gamma}$ | 45 | 68 |
| $T_c/T_F$ | ------ | 0.0013 |
| $k_F$ (10$^6$ cm$^{-1}$) | 3.6 | 8.5 |
| $m^*$ | 0.11 m$_e$ | 0.13 m$_e$ |
| $v_F$ (10$^7$ cm s$^{-1}$) | 3.78 | 7.79 |
| $l$ (nm) | 89.9 | 4.5 |

Figure 4(c) presents the Nernst coefficient measured in presence of magnetic field $H = 8$ T for both undoped Bi$_2$Se$_3$ and Nb$_{0.20}$Bi$_2$Se$_3$. We have observed a large Nernst coefficient ($v = S_{xy}/H$) ~ 2.3 μV K$^{-1}$T$^{-1}$ at room temperature for pristine Bi$_2$Se$_3$, which decreases upon Nb doping (~ 0.5 μV/K$^2$). The large value of Nernst coefficient is usually discussed in literature due to the presence of Dirac dispersion bands at the Fermi level, multiple bands, and fluctuation forms of spin-density-wave (SDW) in iron-pnictide superconducting compounds.[50] Since, Bi$_2$Se$_3$ and its derivatives have Dirac cone at the Fermi surface, large value of Nernst coefficient is consistent with the linear dispersion relation of the bands crossing the Fermi level. The Nernst coefficient in parent Bi$_2$Se$_3$ and Nb$_{0.20}$Bi$_2$Se$_3$ is positive above T = 170 K, with a sign change at 170 K, and reaches a minimum (valley) at ~ 38 K, then finally goes close to zero. Such a sign change observed in Nernst coefficient is not accompanied in Seebeck coefficient, thus ruling out the possibility of multi-band effects. Similar kind of sign change in Nernst coefficient has been reported in earlier in iron pnictide compounds[51] due to the SDW fluctuations, Fe-based superconductors[52] described to the Fermi surface reconstruction and recently in Fe$_3$Sn$_2$[53], that has been attributed to the anomalous Nernst effect. Within the Boltzmann theory, in a single band picture, following expression links the Nernst coefficient ($v$) with the Fermi temperature ($T_F$)[54], $v = \frac{N}{H} = \frac{\pi^2}{3}\frac{k_B}{e}\frac{T}{T_F}\mu$, where $k_B$ is Boltzmann's constant, $e$ is electron charge and $\mu$ is mobility. This formula positively correlates the Nernst effect with the carrier mobility. Thus, undoped Bi$_2$Se$_3$ has more Nernst signal value with the high carrier mobility (~3844 cm$^2$V$^{-1}$s$^{-1}$ at 5 K) than the Nb$_{0.20}$Bi$_2$Se$_3$ which has low carrier mobility (~82 cm$^2$V$^{-1}$s$^{-1}$ at 5 K). The $v/T$ for undoped Bi$_2$Se$_3$ and Nb$_{0.20}$Bi$_2$Se$_3$ in zero-temperature limit is -0.01 μV K$^{-2}$ T$^{-1}$ and -0.0023 μV K$^{-2}$ T$^{-1}$, respectively, which is consistent with the mobility values. Obtained value of the low temperature Nernst signal for Bi$_2$Se$_3$ is in the same range as reported by Fauqué et al.[55] although it depends upon the carrier density. Table I summarizes the parameters obtained from the $S/T$ zero-temperature limit, $S^m/T$ and $v^m/T$ are values obtained at lowest temperature (2 K).

**Conclusion:**

We have investigated the magneto-transport and thermoelectric studies for Nb$_{0.20}$Bi$_2$Se$_3$ single crystals. Our results show the anisotropic superconductivity in angular dependence of resistivity and upper critical field in SC state, which are explained through the anisotropic effective mass Ginzburg-Landau theory. The Hall effect measurements strongly suggest the absence of spontaneous magnetization in superconducting state, which is also supported via the *M-H* plot, thus contradicting the claims of chiral superconducting phase in Nb$_{0.20}$Bi$_2$Se$_3$. The results presenting thermoelectric response on Nb$_{0.20}$Bi$_2$Se$_3$ are compared with pristine Bi$_2$Se$_3$, showing enhancement in Fermi temperature (~ 4 times), with decrease in Seebeck value. The large Nernst coefficient at room temperature is ascribed due to the linearly band dispersion relation in both Bi$_2$Se$_3$ and Nb$_{0.2}$Bi$_2$Se$_3$.

**Acknowledgement:**

We acknowledge IIT Mandi and AMRC, IIT Mandi for the financial support and experimental facility. SS acknowledges MHRD for the HTRA fellowship.

# Supplementary information

**Materials and Methods:**

Single crystals of $Nb_{0.20}Bi_2Se_3$ were synthesised using melt grown technique. High purity elements Bi granular (≥ 99.99 %), Se pellets (≥ 99.999 %) and Nb wire (1 mm dia, 99.8 %) were accurately weighed according to the stoichiometric amount, vacuum sealed and subjected to 3-step heat treatments for homogeneity of the compounds. In first step, $Bi_2Se_3$ was prepared at 850°C for 48 hrs, slowly cooled to 550°C and annealed for 72 hrs, after that it was furnace off cooled to room temperature. In second step, prepared $Bi_2Se_3$ was ground using mortar and pestle and taken in form of pellets, with Nb wire which is subjected to heat treatment at 850°C for 96 hrs, slowly cooled to 620°C for 24 hrs and furnace off cooled to room temperature. In third step, the ingots obtained in previous step were reground, pelletized, and placed to similar heat treatment as in step 2. The single crystals were obtained from ingots by easily cleaving along the basal plane.

Phase purity and crystal structure were studied through X-ray diffraction pattern using a rotating anode Rigaku Smartlab diffractometer in Bragg Brentano geometry (Cu-Kα, λ= 1.5418 Å) on powered as well as thin flake sample. Room temperature Raman measurements were performed using a Horiba LabRAM HR Evolution Raman spectrometer with solid state laser (532 nm). Angular magneto-transport measurements were carried out on a horizontal rotator in PPMS (DynaCool, *Quantum Design Inc.*) in temperature range 1.8 - 300 K and magnetic fields up to 8 T. The standard four-probe method was used for resistivity and Hall effect measurements. The magnetisation measurements were performed using a SQUID magnetometer (MPMS-3, *Quantum Design Inc.*). Thermal transport (Seebeck and Nernst coefficient) measurements were performed on a homebuilt setup integrated with PPMS using multifunctional probe assembly.[1]

**Structural aspects:**

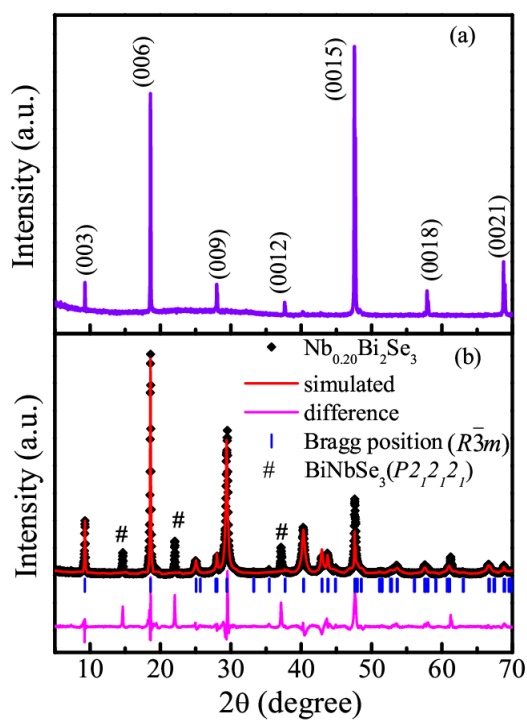

Figure S1: (a) X-ray diffraction pattern of $Nb_{0.20}Bi_2Se_3$ single crystal showing orientation (00*l*) planes, and (b) Rietveld refined powdered XRD pattern.

The X-ray diffraction pattern obtained on flakes reveals single crystalline quality of the sample grown, showing only (00*l*) reflections of the rhombohedral phase as shown in fig. S1(a). The Rietveld refinement (Fig.S1(b)) of the powdered XRD pattern (using FullProf Suite) confirmed the rhombohedral phase (space group: $R\bar{3}m$, #166). The XRD pattern reveal the presence of an impurity phase, $BiNbSe_3$ (space group $P2_12_12_1$, #19), which has been observed in all $Nb_xBi_2Se_3$ compounds reported so far.[2]

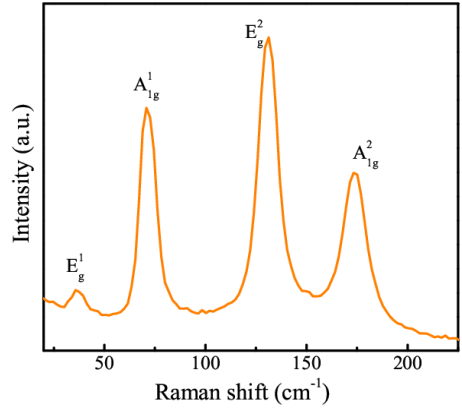

Figure S2: Raman spectra showing the prominent Raman modes.

The Raman spectra (fig. S2) shows the crystalline nature of the material, where four intrinsic Raman active modes viz. $E_g^1$ (35.8 cm$^{-1}$), $A_{1g}^1$ (71.3 cm$^{-1}$), $E_g^2$ (131.4 cm$^{-1}$), $A_{1g}^2$ (174.1 cm$^{-1}$) were clearly observed. The bulk optical phonon modes $E_g$ and $A_{1g}$ corresponds to in-plane and out-of-plane atomic vibrations, respectively. These are well in agreement with the previous reports on Bi$_2$Se$_3$.[3,5]

**Magnetisation studies:**

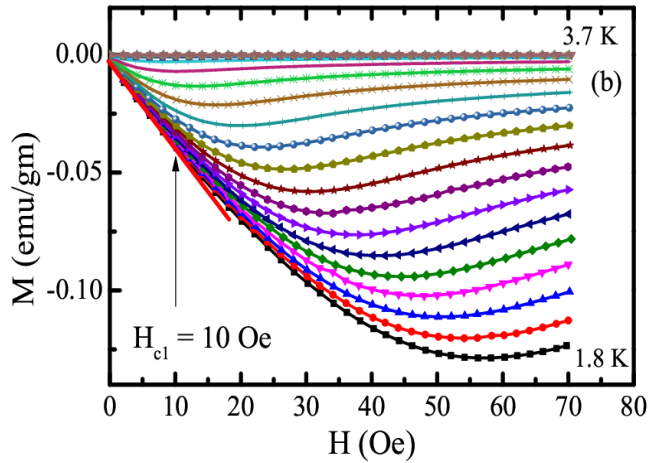

Figure S3 *M-H* plot at different temperature below $T_c$ for extracting $H_{c1}$.

Figure 9 (b) show the field dependent magnetisation measurements performed at various temperatures from 1.8 K to 3.7 K with field sweeps from 0 to 70 Oe, used to determine lower critical field, $H_{c1}$. The lower critical field, $H_{c1}$ (H // *ab*-plane) at 1.8 K ~ 10.23 Oe, which was estimated from the point of deviation of linearity from the magnetisation plot (*M* vs *H*). This apparent value is corrected for the demagnetization effect, using the approximation for the slab geometry[4], given by $H_{c1}^*(0) = \frac{H_{c1}(0)}{\tanh\sqrt{\frac{0.36b}{a}}} = 36.37$ Oe, where *b* (0.52 mm) is thickness and *a* (2.24 mm) is width of the sample.